\documentclass[letterpaper, 10 pt, conference]{ieeeconf}
\IEEEoverridecommandlockouts
\usepackage{acro}
\let\labelindent\relax
\usepackage{enumitem}

\usepackage{graphics}
\usepackage{graphicx} 
\usepackage{multirow}
\usepackage{longtable}
\usepackage{epsfig}
\usepackage{epstopdf}

\usepackage{amsmath, mathrsfs, dsfont}
\usepackage{amssymb}
\usepackage{amsfonts}
\usepackage{bm}

\usepackage{cite}
\usepackage{verbatim}

\usepackage{amscd}
\usepackage{color}
\definecolor{lgray}{gray}{0.6}
\usepackage{rotating}

\usepackage[font=footnotesize]{subcaption}

\usepackage{mathptmx}
\usepackage[11pt]{moresize}
\usepackage{flushend}
\usepackage[stable]{footmisc}

\usepackage{algorithmicx}
\usepackage{algorithm}
\usepackage{algpascal}
\usepackage{float}
\usepackage{algc}
\usepackage{algcompatible}
\usepackage{algpseudocode}

\usepackage[vcentermath]{youngtab} 
\usepackage{url}

\newcommand{\Bolda}				{ \mathbf{a} }

\newcommand{\Boldb}				{ \mathbf{b} }

\newcommand{\Boldd}				{ \mathbf{d} }

\newcommand{\Bolde}				{ \mathbf{e} }
\newcommand{\BoldF}				{ \mathbf{F} }
\newcommand{\Boldf}				{ \mathbf{f} }

\newcommand{\BoldG}				{ \mathbf{G} }

\newcommand{\BoldH}				{ \mathbf{H} }
\newcommand{\Boldh}				{ \mathbf{h} }

\newcommand{\BoldJ}				{ \mathbf{J} }

\newcommand{\BoldP}				{ \mathbf{P} }
\newcommand{\Boldp}				{ \mathbf{p} }
\newcommand{\BoldQ}				{ \mathbf{Q} }
\newcommand{\Boldq}				{ \mathbf{q} }
\newcommand{\BoldR}				{ \mathbf{R} }

\newcommand{\BoldS}				{ \mathbf{S} }

\newcommand{\Boldv}				{ \mathbf{v} }
\newcommand{\Boldw}				{ \mathbf{w} }

\newcommand{\Boldx}				{ \mathbf{x} }

\newcommand{\Boldy}				{ \mathbf{y} }

\newcommand{\0}					{ \boldsymbol{0} }
\newcommand{\1}					{ \boldsymbol{1} }

\newcommand{\Boldomega}			{ \boldsymbol{\omega} }

\newcommand{\BoldPhi}			{ \boldsymbol{\Phi} }

\newcommand{\BoldDelta}			{ \boldsymbol{\Delta} }

\newcommand{\BoldSigma}			{ \boldsymbol{\Sigma} }

\newcommand{\Boldeta}			{ \boldsymbol{\eta} }

\newcommand{\BoldPsi}           {\boldsymbol{\Psi}}

\newcommand\norm[1]{\lVert#1\rVert}

\newcommand\T{\rule{0pt}{2.6ex}}       
\newcommand\B{\rule[-1.2ex]{0pt}{0pt}} 

\newcounter{inlineenum}
\renewcommand{\theinlineenum}{\alph{inlineenum}}

\newcommand{\brmrk}[1]{\begin{remark} \label{#1} }
	\newcommand{\ermrk}{ \hfill $\bigtriangleup$    \end{remark} \vspace{1mm} }

\newtheorem{exercise}{Exercise}[section]
\newcommand{\boex}[1]{\begin{example} \label{#1} --- \rm}
	\newcommand{\eoex}{ \hfill $\bigtriangleup$    \end{example} \vspace{1mm} }
\newtheorem{example}{Example}[section]
\newcommand{\bohw}[1]{\begin{exercise} \label{#1} -- \rm}
	\newcommand{\eohw}{ \hfill    \end{exercise} \vspace{1mm} }
\newtheorem{assumption}{Assumption}[section]
\newcommand{\boass}[1]{\begin{assumption} \label{#1} -- \rm}
	\newcommand{\eoass}{ \hfill    \end{assumption} \vspace{1mm} }

\renewcommand{\baselinestretch}{1.0}

\newcommand{\black}{\color{black}}

\definecolor{brinkpink}{rgb}{1.00, 0.33, 0.64}



\usepackage{tikz}
\usepackage{wrapfig}
\usepackage{stfloats}
\DeclareMathAlphabet\mathbfcal{OMS}{cmsy}{b}{n}
\DeclareAcronym{APC}{
	short=APC,
	long=Antenna Phase Center,
}
\DeclareAcronym{BNB}{
	short=BNB,
	long=branch-and-bound,
}
\DeclareAcronym{COM}{
short=COM,
long=Center of Mass,
}

\DeclareAcronym{CV}{
short=CV,
long=Connected Vehicle,
}

\DeclareAcronym{CAV}{
	short=CAV,
	long=Connected Automated Vehicles,
}

\DeclareAcronym{DCB}{
short=DCB,
long=Differential Code Bias,
}
\DeclareAcronym{DFS}{
	short=DFS,
	long=Depth-First Search,
}
\DeclareAcronym{DSRC}{
	short=DSRC,
	long=Dedicated Short-Range Communications,
}

\DeclareAcronym{CDMA}{
	short=CDMA,
	long=Code Division Multiple Access,
}

\DeclareAcronym{CDF}{
	short=CDF,
	long=Cumulative Distribution Function,
}

\DeclareAcronym{CE-CERT}{
	short=CE-CERT,
	long=College of Engineering Center for Environmental Research and Technology,
}

\DeclareAcronym{CME}{
	short=CME,
	long=common-mode errors,
}

\DeclareAcronym{Full-RAPS}{
	short=Full-RAPS,
	long=RAPS Optimization using Full Information matrix,
}
\DeclareAcronym{Diag-RAPS}{
	short=Diag-RAPS,
	long=RAPS Optimization using Diagonal Information matrix,
}

\DeclareAcronym{FDMA}{
	short=FDMA,
	long=Frequency Division Multiple Access,
}
\DeclareAcronym{DGPS}{
	short=DGPS,
	long=Differential Global Positioning System,
}
\DeclareAcronym{DLR}{
	short=DLR,
	long=German Aerospace Center,
}
\DeclareAcronym{DGNSS}{
short=DGNSS,
long=Differential GNSS,
}
\DeclareAcronym{ECEF}{
short=ECEF,
long=Earth-Centered Earth-Fixed,
}
\DeclareAcronym{ECI}{
	short=ECI,
	long=Earth-Centered Inertial,
}
\DeclareAcronym{EKF}{
	short=EKF,
	long=Extended Kalman Filter,
}
\DeclareAcronym{KF}{
	short=KF,
	long=Kalman Filter,
}
\DeclareAcronym{GPS}{
	short=GPS,
	long=Global Positioning System,
}
\DeclareAcronym{GNSS}{
short=GNSS,
long=Global Navigation Satellite Systems,
}
\DeclareAcronym{GPSSPS}{
short=GPS SPS,
long=GPS standard positioning service,
}

\DeclareAcronym{HMM}
{short = HMM, long = Hidden Markov Model}

\DeclareAcronym{HD}
{short = HD, long = Horizontal Distance}

\DeclareAcronym{HiD}{
	short=Hi-Def,
	long=High-definition,
}

\DeclareAcronym{IMU}{
short=IMU,
long=Inertial Measurement Unit,
}

\DeclareAcronym{ILP}{
short=ILP,
long=Integer Linear Programming,
}

\DeclareAcronym{I2V}{
short=I2V,
long=Infrastructure-to-Vehicle,
}
\DeclareAcronym{IGS}{
	short=IGS,
	long=International GNSS Service,
}
\DeclareAcronym{ISB}{
	short=ISB,
	long=inter-system bias,
}

\DeclareAcronym{LLMM}
{short = LLMM, long = Lane-level Map-matching}

\DeclareAcronym{LMI}
{short = LMI, long = Linear Matrix Inequality}

\DeclareAcronym{LD}
{short = LD, long = Lane Determination}

\DeclareAcronym{LOS}
{short = LOS, long = line-of-sight}

\DeclareAcronym{RLMM}
{short = RLMM, long = Road-level Map-matching}

\DeclareAcronym{MSE}
{short = MSE, long = Mean Square Error}

\DeclareAcronym{MGEX}
{short = MGEX, long = Multi-GNSS Experiment}

\DeclareAcronym{MAP}
{short = MAP, long = maximum-a-posteriori}
\DeclareAcronym{OS}{
	short=OS,
	long=Open Service,
}

\DeclareAcronym{OSB}{
short=OSB,
long=Observable-specific Code Biases,
}
\DeclareAcronym{OSR}{
short=OSR,
long=Observation Space Representation,
}
\DeclareAcronym{PPP}{
	short=PPP,
	long=Precise Point Positioning,
} 

\DeclareAcronym{RTPPP}{
	short=RT-PPP,
	long=Real-time PPP,
}

\DeclareAcronym{PPP-AR}{
short=PPP-AR,
long=Precise Point Positioning Ambiguity Resolution,
} 

\DeclareAcronym{PP}{
short=PP,
long=Post Processing,
}

\DeclareAcronym{PVA}{
	short=PVA,
	long={position, velocity, acceleration},
} 

\DeclareAcronym{RTCM}{
short=RTCM,
long=Radio Technical Commission for Maritime Services,
}
\DeclareAcronym{RTK}{
short=RTK,
long=Real-time Kinematic Positioning,
}
\DeclareAcronym{SBAS}{
short=SBAS,
long=Satellite Based Augmentation Systems,
}

\DeclareAcronym{SNR}{
short=SNR,
long=Signal-to-Noise Ratio,
}
\DeclareAcronym{SSR}{
short=SSR,
long=State Space Representation,
}
\DeclareAcronym{SPS}{
	short=SPS,
	long=Standard Positioning Service,
}

\DeclareAcronym{SAE}{
	short=SAE,
	long=Society of Automotive Engineers,
}
\DeclareAcronym{STEC}{
	short=STEC,
	long=Slant Total Electron Content,
}
\DeclareAcronym{VTEC}{
	short=VTEC,
	long=Vertical Total Electron Content,
}

\DeclareAcronym{SH}{
	short=SH,
	long=Spherical Harmonic,
}
\DeclareAcronym{TGD}{
short=TGD,
long=Timing Goup Delay,
}
\DeclareAcronym{TD}{
	short=TD,
	long=Threshold Decisions,
}
\DeclareAcronym{TOP}{
	short=TOP,
	long=time-of-signal-propagation,
}
\DeclareAcronym{TOT}{
	short=TOT,
	long=time-of-signal-transmission,
}
\DeclareAcronym{TOR}{
	short=TOR,
	long=time-of-signal-reception,
}
\DeclareAcronym{ZTD}{
short=ZTD,
long=Zenith Troposphere Delay,
}
\DeclareAcronym{TEC}{
short=TEC,
long=Total Electron Content,
}
\DeclareAcronym{IPP}{
	short=IPP,
	long=Ionosphere Pierce Point,
}
\DeclareAcronym{NOAA}{
	short=NOAA,
	long=National Oceanic and Atmospheric Administration,
}
\DeclareAcronym{UCR}{
	short=UCR,
	long=University of California-Riverside,
}
\DeclareAcronym{USTEC}{
short=US-TEC,
long=US Total Electron Content,
}
\DeclareAcronym{VNDGNSS}{
	short=VN-DGNSS,
	long=Virtual Network DGNSS,
}
\DeclareAcronym{IOD}{
	short=IOD,
	long=Issue Of Data,
}
\DeclareAcronym{CAS}{
	short=CAS,
	long=Chinese Academy of Sciences,
}
\DeclareAcronym{CNES}{
	short=CNES,
	long=Centre National d'Études Spatiales,
}
\DeclareAcronym{RMS}{
	short=RMS,
	long=Root Mean Square,
}
\DeclareAcronym{SF}{
	short=SF,
	long=Single Frequency,
}
\DeclareAcronym{STD}{
	short=STD,
	long=Standard Deviation,
}
\DeclareAcronym{DF}{
	short=DF,
	long=Dual Frequency,
}
\DeclareAcronym{ICD}{
	short=ICD,
	long=Interface Control Document,
}
\DeclareAcronym{NED}{
	short=NED,
	long={North, East and Down},
}
\DeclareAcronym{WAAS}{
	short=WAAS,
	long=Wide Area Augmentation System,
}
\DeclareAcronym{OPUS}{
	short=OPUS,
	long=Online Positioning User Service,
}
\DeclareAcronym{PDF}{
	short=PDF,
	long=Probability Density Function,
}
\DeclareAcronym{RAPS}{
	short=RAPS,
	long=Risk-Averse Performance-Specified,
}

\DeclareAcronym{VRS}{
	short=VRS,
	long=Virtual Reference Station,
}

\DeclareAcronym{SPP}{
	short=SPP,
	long=Single-frequency Point Positioning,
}

\DeclareAcronym{SPaT}{
	short=SPaT,
	long=Signal Phase and Timing,
}

\DeclareAcronym{TECU}{
	short=TECU,
	long=Total Electron Content Units,
}

\DeclareAcronym{BNC}{
	short=BNC,
	long=BKG NTRIP Client,
}

\DeclareAcronym{ITS}{
	short=ITS,
	long=Intelligent Transportation Systems
}

\DeclareAcronym{USDOT}{
	short=USDOT,
	long=U.S. Department of Transportation
}

\DeclareAcronym{WHU}{
	short=WHU,
	long=Wuhan University
}

\begin{document}
	
	\title{Convex Reformulation of  Information Constrained  Linear State Estimation with Mixed-Binary Variables for Outlier Accommodation}
	\author{Wang~Hu,~Zeyi~Jiang,~Hamed~Mohsenian-Rad,
		and~Jay~A.~Farrell
		\thanks{W. Hu, Z. Jiang, H. Mohsenian-Rad and J. A. Farrell are with the Department of Electrical and Computer Engineering,
			University of California, Riverside, CA 92521, USA. (e-mail: \{whu027,~zjian037\}@ucr.edu, \{hamed,~farrell\}@ece.ucr.edu).}
	}
	\maketitle
	
\begin{abstract}
This article considers the challenge of accommodating outlier measurements in state estimation. 
The \ac{RAPS} state estimation approach addresses outliers as a measurement selection Bayesian risk minimization problem subject to an  information  accuracy constraint, which is a non-convex optimization problem.  
Prior explorations into \ac{RAPS} rely on exhaustive search, which becomes computationally infeasible as the number of measurements is increases. 
This paper derives a convex formulation for the \ac{RAPS} optimization problems via transforming the mixed-binary variables into linear constraints. 
The convex reformulation herein can be solved by convex programming toolboxes, significantly enhancing computational efficiency. 
We explore two specifications: Full-RAPS, utilizing the full information matrix, and Diag-RAPS, focusing on diagonal elements only.
The simulation comparison demonstrates that Diag-RAPS is faster and more efficient than Full-RAPS.
In comparison with  \ac{KF} and \ac{TD}, Diag-RAPS consistently achieves the lowest risk, while achieving the performance specification when it is feasible.
\end{abstract} 	
		
\section{Introduction}	
The rapid decrease in the cost and size of sensors has led to an abundant number of measurements, far exceeding the number required for state observability.
For instance, position determination using \ac{GNSS}  is possible with pseudorange measurements from four spatially diverse satellites \cite{misra2006global}.
There are now four globally operational satellite constellations (e.g., BeiDou, GPS, Galileo, GLONASS) each providing pseudorange measurements on multi-frequencies.
The number of GNSS measurements per epoch, under open-sky conditions, is predicted to increase toward 100 in the near future \cite{camacho2017current,odolinski2024evaluation}.
Similarly,  feature-based navigation only requires four features with diverse directions for observability  \cite{ramanandan2010observability,9867337,song_deep_2019,kang20216},
while cameras can provide hundreds of features per image \cite{li-19-segmentation}.
\black
Such applications, where the number of available measurements far exceeds the number required for observability are {\em signal-rich}.

For signal-rich applications, there is an important performance versus Bayesian risk trade-off.
If all measurements were outlier-free with known accuracy, then using all measurements within a maximum a posteriori framework would yield the optimal state estimation performance \cite{deng2024incremental,deng2023long}. 
However, in typical applications
some measurements are affected by outliers.
Therefore, each measurement that is used has the potential to provide enhanced estimation accuracy as quantified through the state error covariance matrix, if the measurement is outlier-free.
Alternatively, there is the risk that outlier affected measurements will corrupt the state estimate while causing the estimator to be overly confident in that corrupted estimate (i.e., the error covariance matrix is too small).

This trade-off and methods to directly address it are critically important because
 at any measurement epoch several measurements may contain outliers.
When the number of measurements that are used is large, adding one additional valid measurement increases the accuracy only marginally (i.e., covariance decreases proportional to $\frac{1}{m}$, where $m$ is the number of measurements used); however, use of a single outlier-affected measurement can destroy the validity of all subsequent information processing by corrupting both the mean and covariance of the state estimate.

State estimation using discrete-time measurements involves two steps.
The time update propagates the posterior state estimate from the previous measurement epoch across the time span between measurements, to provide a prior state estimate at the next measurement epoch.
The measurement update incorporates the measurement information to correct the prior state estimate to produce a posterior state estimate at the measurement time. 
Because outliers only affect the measurement update,
this paper focuses entirely on the problem of measurement selection for the measurement update, with a focus on signal-rich environments.

Outliers are the measurements that are inconsistent with the model and the remainder of the set of measurements  \cite{barnett1994outliers}. 
Accommodation of outliers in state estimation is a well-studied problem.
Standard methods for state estimation and control (see e.g.: \cite{frank1997survey,patton1994review}) evaluate measurement residuals against their expected values using a fixed or an adaptive threshold test, as will be reviewed in Section \ref{subsect:TD}. 
More advanced outlier accommodation methods have been presented in other applications \cite{ rousseeuw2006computing, maronna2019robust,zhu2024ensemble,li-23-deception-detection}, such as Least Squares-based linear model fitting. 
These methods either detect and ignore outliers or de-weight measurements with large residuals.
None of these approaches consider the risk-versus-performance trade-off involved with all measurements.

\ac{RAPS} is an alternative approach to addressing outliers in state estimation \cite{hu2024rapsppp,hu2024rtk,Aghapour_TCST_2019}.
This approach, reviewed in Section \ref{subsect:RAPS}, 
selects measurements to minimize a cost function that quantifies Bayesian risk
while satisfying a performance constraint specified through the information matrix. 
Prior work on \ac{RAPS} presented the optimization problem in its natural non-convex form.  
To achieve the global optimal solution, an \ac{DFS} approach was suggested to examine all $2^m$ combinations of the binary measurement selection vector \cite{zhang2024development}. \black 
This strategy becomes computationally infeasible as $m$ increases. 
This article presents and demonstrates a convex reformulation of the \ac{RAPS} optimization problems by converting transforming the mixed-binary variables into linear constraints, enabling efficient solution using modern optimization tools. 
We explore and evaluate two approaches to specifying performance: one involving the full information matrix and another focusing solely on its diagonal elements. 
Additionally, this paper includes a simulation comparison of the information and risk outcomes from \ac{KF} using all measurement, \ac{TD} for outlier removal, and \ac{RAPS} for measurement selection.
\black

\section{Problem Statement}\label{sect:problem_statement}
The discrete-time evolution of the linear system is   
\begin{align} \label{eqn:state_prop_model}
	\Boldx_{k} &= \BoldF \, \Boldx_{k-1} +  \Boldw_k 
\end{align}
where $\Boldx_k \in\mathbb{R}^n$ is the state vector at discrete-time $t_k = k\,T$ where $T$ is the sampling interval, 
$\BoldF$ is the known transition matrix and $\Boldw_k \sim \mathcal{N}(\0,\,\BoldQ)$ is white Gaussian process noise with covariance matrix $\BoldQ$.
The time propagation of the state vector $\Boldx_k$ and its covariance $\BoldP_{k}$ is assumed to  be unaffected by outliers. 
Given the Gaussian probability density function of the posterior $\mathcal{N}(\hat{\Boldx}_{k-1}^+,\,\BoldP_{k-1}^+)$, the state estimate  and its error covariance are propagated through time as
\begin{eqnarray}
	\hat{\Boldx}_{k}^- &\doteq& \BoldF \, \hat{\Boldx}_{k-1}^+ \label{eqn:state_tm_update}\\
	\BoldP_{k}^- &\doteq& \BoldF \, \BoldP_{k-1}^+   \, \BoldF^\top +   \BoldQ  \label{eqn:cov_tm_update}
\end{eqnarray}

The measurement vector at time $t_k$ is modeled as:
\begin{align} \label{eqn:MeasurementModel}
	\Boldy_k = \BoldH \, \Boldx_k + \Boldeta_k 
\end{align}
where $\Boldy_k\in \mathbb{R}^m$, $\BoldH \in \mathbb{R}^{m \times n}$ is the known measurement matrix, and  $\Boldeta_k \sim \mathcal{N}(\0,\,\BoldR_k)$ represents the  white Gaussian measurement noise.
The measurement noise matrix $\BoldR_k\in \mathbb{R}^{m \times m}$  is assumed to be diagonal and invertible. 
Therefore, it can be written as $\BoldR = \sum_{i=1}^{m} \sigma_{i}^2 \, \Bolde_i  \,{\Bolde_i}^\top$, where $\Bolde_i$ is the $i^{th}$ standard basis vector.

Some measurements might be affected by outliers.
Outlier measurements are those that are improbable relative to the model stated in eqn. (\ref{eqn:MeasurementModel}).
They may described as
\begin{equation}\label{eqn:MeasurementModel_s}
	y_k(i) = \Boldh_i \, \Boldx_k + \eta_k(i) +s_i,
\end{equation}
where $y_k(i)$ represents the $i$-th element of $\Boldy_k$, that is affected by the outlier $s_i$, $\eta_k(i)$ is the $i$-th element of $\Boldeta_k$, and $\Boldh_i$ is the $i$-th row of $\BoldH$. 
The outlier $s_i$ is unknown and has no established model.
The challenge lies in accommodating these outlier measurements during the measurement update process.

\section{State Estimation Using Selected Measurements} \label{sect:MAP}
This section reviews different strategies to accommodate outliers in the state estimation measurement update.
Each strategy either selects or discards certain measurements. 
The discussion is facilitated by introducing a binary vector $\Boldb = [b_1,\,...\,,b_m]^\top$.
The binary element $b_i$ indicates whether to include ($b_i=1$) or exclude ($b_i=0$) measurement $y_k(i)$ \cite{carlone2014selecting}. 

For a given value of $\Boldb$, the \ac{MAP} approach estimates the state by minimizing the cost function
\begin{align}\label{eqn:state_estimate}
	\hat{\Boldx}_k^+ &= \underset{\Boldx_k}{\text{argmin}}~C(\Boldx_k,\,\Boldb),
\end{align}
where the cost function (see \cite{hu2024rapsppp}) is
\begin{align}
	C(\Boldx_k,\,\Boldb) &= \norm{ \Boldx_k - \hat{\Boldx}^-_k}_{\BoldP_k^-}^2 + \left\|\BoldSigma_{\BoldR} \, \BoldPhi(\Boldb)\left(\Boldy_k - \BoldH\,\Boldx_k\right) \right\|^2 \nonumber \\
	&= \norm{ \Boldx_k - \hat{\Boldx}^-_k}_{\BoldP_k^-}^2 + \sum_{i=1}^{m} \frac{b^2_i}{\sigma_i^2} (y_i - \Boldh_i \Boldx)^2 \label{eqn:Costb2} \\
	&= \norm{ \Boldx_k - \hat{\Boldx}^-_k}_{\BoldP_k^-}^2 + \sum_{i=1}^{m} \frac{b_i}{\sigma_i^2} (y_i - \Boldh_i \Boldx)^2 \label{eqn:Costfcn}
\end{align}
where 
$\forall \, \Boldf \in \Re^\ell,\,\BoldPhi(\Boldf) = diag(\Boldf):  \Re^\ell \mapsto \Re^{\ell\times \ell}$ and $b_i^2 = b_i$ due to the binary nature of $b_i$. The covariance matrices $\BoldR^{-1} = {\BoldSigma_{\BoldR}}^\top \, \BoldSigma_{\BoldR} $ and $(\BoldP_k^-)^{-1} =  {\BoldSigma_{\BoldP}}^\top \, \BoldSigma_{\BoldP}$ are represented using the squared Mahalanobis distance (e.g.,
$\norm{ \Boldx_k - \hat{\Boldx}^-_k}_{\BoldP_k^-}^2= \norm{ \BoldSigma_{\BoldP}\,(\Boldx_k - \hat{\Boldx}^-_k)}^2$ see \cite{dellaert_square_2006}).

The posterior information matrix is computed as
\begin{align}
	\BoldJ_k^+ &= \BoldH^\top \BoldPhi(\Boldb)^\top \BoldR_k^{-1}\BoldPhi(\Boldb)\BoldH
	+  \BoldJ_k^- \nonumber \\
	&= \sum_{i=1}^{m} \frac{b_i}{\sigma^2_i} \Boldh^\top_i \Boldh_i + \BoldJ_k^-  \label{eqn:post_info}
\end{align}
where $\BoldJ_k^- = (\BoldP_k^-)^{-1}$ denotes the prior information matrix. 
The corresponding posterior covariance matrix is computed as $\BoldP^+_k = (\BoldJ^+_k)^{-1}$. 
Two approaches to calculate the measurement selection vector $\Boldb$ are considered below.

	\subsection{Threshold Decisions (TD)} \label{subsect:TD}
	The traditional method utilizes a threshold test to decide whether a measurement is compromised by an outlier:
\begin{equation} \label{eqn:threshold_test}
	b_{i} \doteq \left\{
	\begin{array}{ll}
		0, & \mbox{ when }{|r_i|} \ge \lambda\sigma_{r_i}\\
		1, & \mbox{ when }{|r_i|} <   \lambda\sigma_{r_i}
	\end{array}
	\right.
\end{equation}
where $\lambda>0$ is the decision threshold, 
$\sigma_{r_i}^2= {\Boldh_i \, \BoldP^- \, \Boldh_i^\top + \sigma_{i}^2}$ 
is the covariance of residual component $r_i$ (see e.g.: \cite{frank1997survey,patton1994review}).
The residuals are computed as
\begin{equation} \label{eqn:residual} 
	r_i \doteq  y_i - \Boldh_i \, \hat{\Boldx}_k^-. 
\end{equation}
After $\Boldb_{TD}$ is determined by eqn. \eqref{eqn:threshold_test}, the optimal estimate and information are the solutions of eqn. \eqref{eqn:state_estimate} and eqn. \eqref{eqn:post_info}.

Many studies have critiqued the \ac{TD} method  for its failure to reliably detect outliers, pointing out its susceptibility to missed detections and false alarms \cite{hu2024rapsppp,aghapour2019outlier}. 
Missed detections can result in corrupted state estimates and overconfident covariance matrices, creating inconsistencies between actual and estimated uncertainties. 
This discrepancy can significantly impair the reliability of future subsequent outlier decisions.
In addition, fixed decision threshold means that outlier decisions  do not take into account any  performance specifications. 
\black

	\subsection{Risk-Averse Optimization}  \label{subsect:RAPS}
	The \ac{RAPS} approach  selects an optimal subset of measurements that to minimize the Bayesian risk as quantified by eqn. \eqref{eqn:Costfcn}, subject to a specified performance criteria  \cite{hu2024rapsppp,Aghapour_TCST_2019}. 
Two constraint implementations are considered herein. 

To facilitate the discussion, we define two symbols.
Dropping the subscript $k$, starting from eqn. \eqref{eqn:post_info} the posterior information matrix computed for a specific choice of $\Boldb$ is
\begin{align}
	\BoldJ^+ (b) &=  \sum_{i=1}^{m} \frac{b_i}{\sigma^2_i} \Boldh^\top_i \Boldh_i + \BoldJ^-  .
	\label{eqn:defn_J_b}
\end{align}

\subsubsection{Full-RAPS}\label{sect:fullraps}
This approach constrains the posterior information matrix within a semi-definite framework: $\BoldJ^+ (b) \succcurlyeq \BoldPhi(\BoldJ_d)$ where the symbol $\BoldJ_d$ represents a user-defined non-negative performance lower bound vector.. 
Defining  $\BoldF_i = \frac{1}{\sigma^2_i} \Boldh^\top_i \Boldh_i$, 
eqn. \eqref{eqn:defn_J_b} becomes
\begin{align}\label{eqn:fullrapsconstraint}
	\sum_{i=1}^{m} \BoldF_i \, b_i \succcurlyeq \BoldPhi(\BoldJ_d) - \BoldJ^-,
\end{align}
With this  \ac{LMI}, the  Full-RAPS  optimization problem is
\begin{equation} \label{eqn:RAPS_full}
	\left.
	\begin{aligned}
		\Boldx_k^+, \Boldb^\star  &= \,\underset{\Boldx_k,\Boldb}{\text{argmin}} ~ C(\Boldx_k,\Boldb)  \\ 
		&\text{s.t.:} \ \ \sum_{i=1}^{m} \BoldF_i \, b_i \succcurlyeq \BoldPhi(\BoldJ_d) - \BoldJ^- \\
		&\ \ \ \ \ b_i \in \{0,\,1\} \ \text{for} \ i = 1,\,...,\,m
	\end{aligned}
	\right\}
\end{equation}
where the cost $C(\Boldx_k,\Boldb)$ is defined in eqn. \eqref{eqn:Costfcn}.

\subsubsection{Diag-RAPS}\label{sect:diagraps}
This approach chooses the constraint $diag(\BoldJ^+ (b)) \geq \BoldJ_d$.
By constraining only the diagonal elements of the posterior information matrix computational complexity  is reduced relative to the semi-definiteness check in eqn. \eqref{eqn:fullrapsconstraint}.
Appendix A of \cite{hu2024optimization}, shows this constraint is equivalent to
\begin{equation} \label{eqn:linear_constraint}
	\BoldG \, \Boldb \ge \Boldd \mbox{ where }	\Boldd = \BoldJ_d - diag(\BoldJ^-_k)
\end{equation}
and
$
	\BoldG =\begin{bmatrix}
		\frac{h_{11}^2}{\sigma_1^2} & \ldots & \frac{h_{m1}^2\T}{\sigma_m^2\B}  \\
		\vdots & \ddots & \vdots  \\
		\frac{h_{1n}^2}{\sigma_1^2} & \ldots & \frac{h_{mn}^2\T}{\sigma_m^2\B}  \\
	\end{bmatrix}
. 
$
With this linear constraint on $\Boldb$,
the Diag-RAPS optimization problem is
\begin{equation} \label{eqn:RAPS_diag}
	\left.
	\begin{aligned}
		\Boldx_k^+, \Boldb^\star & = \,\underset{\Boldx_k,\Boldb}{\text{argmin}} ~ C(\Boldx_k,\Boldb)  \\ 
		&\text{s.t.:} \ \ \BoldG \, \Boldb \ge \Boldd \\
		&\ \ \ \ \ b_i \in \{0,\,1\} \ \text{for} \ i = 1,\,...,\,m
	\end{aligned}
	\right\}
\end{equation}

\subsubsection{Comparison of Full and Diagonal RAPS}
The cost function of eqn. \eqref{eqn:Costfcn} is quadratic in $\Boldx_k$ and linear in $\Boldb$ when these variables are treated separately; however, it is jointly non-convex. 
Therefore, Problems \eqref{eqn:RAPS_full} and \eqref{eqn:RAPS_diag} are each  mixed-integer non-convex problems. 
Given the binary nature of $\Boldb$,
\cite{Aghapour_TCST_2019} proposed exhaustive search methods with worst-case time complexity $O(m!)$. For each choice of $\Boldb$, the optimization which is quadratic in  $\Boldx_k$ is straightforward.
The search over $\Boldb$ could be implemented using an $m$-layer binary tree search, leveraging the standard \ac{BNB} with worst-case time complexity to $O(2^m)$.
%
\black

\section{Convex Reformulation} \label{sect:cvx}
There now exist software packages to solve
convex mixed-integer programming problems; however, 
Problems \eqref{eqn:RAPS_full} and \eqref{eqn:RAPS_diag} are characterized by non-convex objective functions coupled with convex constraints.
This section transforms the non-convex optimization cost function into a convex format,
thereby transforming Problems \eqref{eqn:RAPS_full} and \eqref{eqn:RAPS_diag} into convex problems. 

The cost function in eqn. \eqref{eqn:Costb2} is equivalent to
\begin{align}
	C(\Boldx_k,\,\Boldb) = \norm{ \Boldx_k - \hat{\Boldx}^-_k}_{\BoldP_k^-}^2 + \sum_{i=1}^{m} \frac{1}{\sigma_i^2} (y_i\,b_i - \Boldh_i\,b_i\,\Boldx)^2 \label{eqn:new_cost}.
\end{align}
Let $\Boldq = [\Boldq_1;\,...;\,\Boldq_i;\,...;\,\Boldq_m] \in \mathbb{R}^{n\times m}$, 
where $\Boldq_i = b_i\,\Boldx = [q_{i1},,...,,q_{ij},,...,,q_{in}]^\top \in \mathbb{R}^{n}$ with $q_{ij} = b_i\,x_j$.
Although the product $b_i\,x_j$ is inherently non-convex and mixed-binary, it can be encapsulated within a linear constraint set:
\begin{equation} \label{eqn:cvx_set}
	\mathcal{A}(\Boldx_k, \Boldb, \Boldq) =
	\left\{
	\begin{aligned}
		&\mathcal{A}_1:\,\Boldx_{min} \leq \Boldx \leq \Boldx_{max},\\
		&\mathcal{A}_2:\,\Boldx - \Boldx_{max} (1-b_i) \leq \Boldq_i \leq b_i\,\Boldx_{max}, \\
		&\mathcal{A}_3:\,b_i\,\Boldx_{min} \leq \Boldq_i \leq \Boldx - \Boldx_{min} (1-b_i),
	\end{aligned}
	\right.
\end{equation}
where $\Boldx_{min}$ and $\Boldx_{max}$ denote  lower  and upper bounds on $\Boldx$, respectively. 
For state estimation problems, the posterior state estimate is typically within a limited distance to the prior state. Therefore, these bounds can be configured as $\hat{\Boldx}^- \pm \BoldDelta$, with $\BoldDelta$ being a suitably large positive value.

The linear constraints in eqn. $\mathcal{A}_2$ and eqn. $\mathcal{A}_3$ establish important properties for $\Boldq$:
\begin{enumerate}
	\item For $b_i = 1$,  $\mathcal{A}_2$ simplifies to $\Boldx \leq \Boldq_i \leq \Boldx_{max}$, and   $\mathcal{A}_3$ simplifies to $\Boldx_{min} \leq \Boldq_i \leq \Boldx$, ensuring $\Boldq_i = \Boldx$.
	\item For $b_i = 0$,  $\mathcal{A}_2$ becomes $\Boldx - \Boldx_{max} \leq \Boldq_i \leq 0$, and   $\mathcal{A}_3$ becomes $0 \leq \Boldq_i \leq \Boldx - \Boldx_{min}$, ensuring $\Boldq_i = \mathbf{0}$.
\end{enumerate}
These linear (convex) constraints therefore ensure  $q_{ij}=b_i\,x_j$.
Note that the binary vector $\Boldb$ cannot be relaxed to a non-binary variable, as that would  compromises the require property that $\Boldq_i=b_i,\Boldx$.

For binary $\Boldb$,  with the constraints in eqns. \eqref{eqn:cvx_set},
the cost function in eqn. \eqref{eqn:Costb2} has the convex form:
\begin{align}
	C_c(\Boldx_k, \Boldb, \Boldq) = \norm{ \Boldx_k - \hat{\Boldx}^-_k}_{\BoldP_k^-}^2
	+   \sum_{i=1}^{m} \frac{(y_i\,b_i - \Boldh_i\,\Boldq_i)^2}{\sigma_i^2}. \nonumber
\end{align}
Both Problems \eqref{eqn:RAPS_full} and  \eqref{eqn:RAPS_diag}) can now be written in convex form.
\ac{Full-RAPS}   is equivalent to
\begin{equation} \label{eqn:RAPS_full_convex}
	\left.
	\begin{aligned}
		\Boldx_k^+, \Boldb^\star,\Boldq &= \,\underset{\Boldx_k,\Boldb,\Boldq}{\text{argmin}} ~ C_c(\Boldx_k, \Boldb, \Boldq) \\ 
		&\text{s.t.:} \ \ \sum_{i=1}^{m} \BoldF_i \, b_i \succcurlyeq \BoldPsi(\BoldJ_d) - \BoldJ^-_k \\
		&\ \ \ \ \ b_i \in \{0,\,1\} \ \text{for} \ i = 1,\,...,\,m \\
		&\ \ \ \ \ \mathcal{A}(\Boldx_k, \Boldb, \Boldq),
	\end{aligned}
	\right\}
\end{equation}
\ac{Diag-RAPS}  is equivalent to 
\begin{equation} \label{eqn:RAPS_diag_convex}
	\left.
	\begin{aligned}
		\Boldx_k^+, \Boldb^\star,\Boldq &= \,\underset{\Boldx_k,\Boldb,\Boldq}{\text{argmin}} ~ C_c(\Boldx_k, \Boldb, \Boldq) \\ 
		&\text{s.t.:} \ \ \BoldG \, \Boldb \ge \Boldd \\
		&\ \ \ \ \ b_i \in \{0,\,1\} \ \text{for} \ i = 1,\,...,\,m \\
		&\ \ \ \ \ \mathcal{A}(\Boldx_k, \Boldb, \Boldq).
	\end{aligned}
	\right\}
\end{equation}
Both Problem \eqref{eqn:RAPS_full_convex} and Problem \eqref{eqn:RAPS_diag_convex} are convex mixed-integer optimization problems featuring nonlinear objective functions. 
Problem \eqref{eqn:RAPS_diag_convex} incorporates linear constraints. In contrast, Problem \eqref{eqn:RAPS_full_convex}, not only has linear constraints, but also involves a \ac{LMI} constraint, adding a layer of complexity to its formulation. Both involve binary variables.\black

\black
Both problems are solvable using the \ac{BNB} algorithm in YALMIP \cite{Lofberg2004}. 
While in the  worst case the \ac{BNB} algorithm exhibits  exponential time complexity, on average it is more efficient than the  previously proposed  DFS strategy. 
In principle, both Problem \eqref{eqn:RAPS_full_convex} and Problem \eqref{eqn:RAPS_diag_convex} could be solved more efficiently by combining the \ac{BNB} algorithm with convex optimization. 
However, contemporary solvers like MOSEK and Gurobi \cite{mosek,gurobi}, 
that are commonly used in the CVX \cite{grant2009cvx} and YALMIP optimization environments, 
do not currently support integer variables with semi-definite matrix inequalities.
Therefore, MOSEK and Gurobi are suitable for Problem \eqref{eqn:RAPS_diag_convex}, but not
Problem \eqref{eqn:RAPS_full_convex}.
A comparative analysis of computation times is presented in Sec. \ref{sec:full_diag}, using YALMIP BNB for \ac{Full-RAPS} and CVX MOSEK for {Diag-RAPS}.
\black
%

\section{Evaluation in Simulation} \label{sect:results}
This section evaluates and compares \ac{Full-RAPS} and \ac{Diag-RAPS} on computation time.
Then, state estimation performance is evaluated for: 
the \ac{KF} without outlier removal, 
the \ac{KF} equipped with \ac{TD} outlier removal as discussed in Section \ref{subsect:TD} with $\lambda = 2$, and \ac{Diag-RAPS} solved using MOSEK.
The performance evaluation was conducted in MATLAB on an AMD 5800X CPU.

\black
\subsection{System Model}
The systems that is simulated has the state vector
$$\Boldx = \begin{bmatrix}
	\Boldp & \Boldv & \Bolda
\end{bmatrix}^\top\in\mathbb{R}^9$$
where
$\Boldp\in\mathbb{R}^3$ represents position (components denoted as $N$, $E$, and $D$), 
$\Boldv\in\mathbb{R}^3$ represents velocity,  and 
$\Bolda\in\mathbb{R}^3$ represents acceleration.
Therefore, $n=9$. 
During the simulation, the acceleration signal is selected to cause the vehicle to follow a trajectory similar to a vehicle driving around a city block that is square with each edge being 200 meters long. 
The origin of the coordinate system is at the center of this square. 
The $D$ component undulates in a slow sinusoidal fashion as the vehicle traces its trajectory.

The navigation system does not know the acceleration of the vehicle. 
For the purpose of state estimation, the differential equations for this system are
\begin{align}
	\dot\Boldp(t) = \Boldv(t),~
	\dot\Boldv(t) = \Bolda(t), \mbox{ and }
	\dot\Bolda(t) = \Boldomega(t) 
\end{align}   
where $\Boldomega(t)$ is continuous-time white Gaussian process noise with power spectral density $\BoldS$. 
This is referred to a position-velocity-acceleration model and is frequently used in \ac{GNSS} applications \cite{rahman2020positionaccuracy,hu2024optimization}.

To estimate the state vector, each state estimator uses a vector of measurements that occurs every $T=1$ second.
Each estimator is designed based on the model in eqn. \eqref{eqn:MeasurementModel}.
Each row of the measurement matrix has the form
\begin{align} \label{eqn:h_def}
	\Boldh_i =
	\begin{bmatrix}
		h_{iN} & h_{iE} & h_{iD} & 0 & 0 & 0 & 0 & 0 & 0	\end{bmatrix},
\end{align}
where $\|\Boldh_i\|_2=1$ and $h_{iz}>0$. 
These first three components, related to the position vector, represent the line-of-sight unit vector pointing from the receiver antenna to the satellite (see Section 8.2 in \cite{farrell2008aided}). 
The simulation considers $m$ satellites, with randomly distributed azimuth and elevation, that are fixed during the duration of the simulation.

\subsection{Measurements and Outliers}
The simulated measurements are produced using the model in eqn. (\ref{eqn:MeasurementModel_s}), where $var(\eta_i)=\sigma_i^2$ with $\sigma_i=1.5$ m. 
The ground truth trajectory and outlier corrupted measurements were created once and then used by all three estimators.

Each measurement outlier signal $s_i(kT)$ was defined to have properties similar to multipath in GNSS. 
Dropping the $i$, because the model is the same for all measurements, the variance of each outlier is 
$$\sigma_s^2 = \sigma_{el}^2 + \sigma_{az}^2,$$
where
$
\sigma_{el} = \frac{0.6}{|\psi| + \epsilon}$
and 
$	\sigma_{az} = \frac{0.3}{\theta + \epsilon}.
$
In this expression, $\theta>0$ and $\psi$ represent the radian elevation and azimuth angles of the satellite relative to the receiver and $\epsilon$ is a small positive number to prevent division by zero. 
For the $i$-th measurement, at each time $t_k=k\,T$, an outlier  $s_i(kT)$ is computed as a zero mean Gaussian random variable with standard deviation $\sigma_{s_i}^2$ based on  $\theta_i$ and $\psi_i$.

Because the number of reflecting surfaces along the signal path that can cause multipath increases as elevation decreases, the outlier variance is designed to increase rapidly as the satellite elevation decreases.
The variance of the outlier with respect to azimuth is designed to correspond to a building or other reflecting surface near the origin of the simulated system that causes high multipath.
Because the vehicle trajectory goes around the building the specific direction corresponding to the high variance changes with the vehicle location.

\subsection{Performance Specification}
The performance specification $\BoldJ_d$, defines a the lower bound in the constraint in Problems \eqref{eqn:RAPS_full_convex} and  \eqref{eqn:RAPS_diag_convex}.
It was selected to only constrain the information required for the three position variables. 
The \ac{SAE} specification for positioning accuracy in highway vehicle applications requires the horizontal accuracy better than 1.5 meters and vertical accuracy 3 meters, both at a 68\% probability level. 
The determination of $\BoldJ_d$ based on the \ac{SAE} criteria is discussed in Section 7.2 of \cite{hu2024optimization}, yielding  $$\BoldJ_d = [1.389,\,1.389,\,0.347].$$

Additional states could be constrained, but this constraint is appropriate for this application. 
If constraints were desired for 
velocity (or acceleration) which are not observable until position measurements are available for at least two (three) distinct times, then the lower bound would need to change with time.

\subsection{Runtime comparison between \ac{Full-RAPS} and \ac{Diag-RAPS}}\label{sec:full_diag}

This section evaluates the real-time feasibility of the  \ac{Full-RAPS} and \ac{Diag-RAPS} approaches.
Four experiments were conducted, each utilizing a fixed number of 15, 20, 25, and 30 measurements per epoch. 
The \ac{Full-RAPS} measurement update corresponds to the solutions of the optimization problems defined in Problem \eqref{eqn:RAPS_full_convex}, solved using the YALMIP BNB toolbox. The \ac{Diag-RAPS} update pertains to the solutions of the optimization problems defined in Problem \eqref{eqn:RAPS_diag_convex}, solved via CVX MOSEK toolbox. The exhaustive search strategy in \cite{Aghapour_TCST_2019} was not included in this study due to the prohibitive computational cost (e.g., more than 10 minutes for a single epoch with $m=15$), especially as $m$ increases.

\begin{figure}[tb]
	\centering
	\includegraphics[width=\columnwidth]{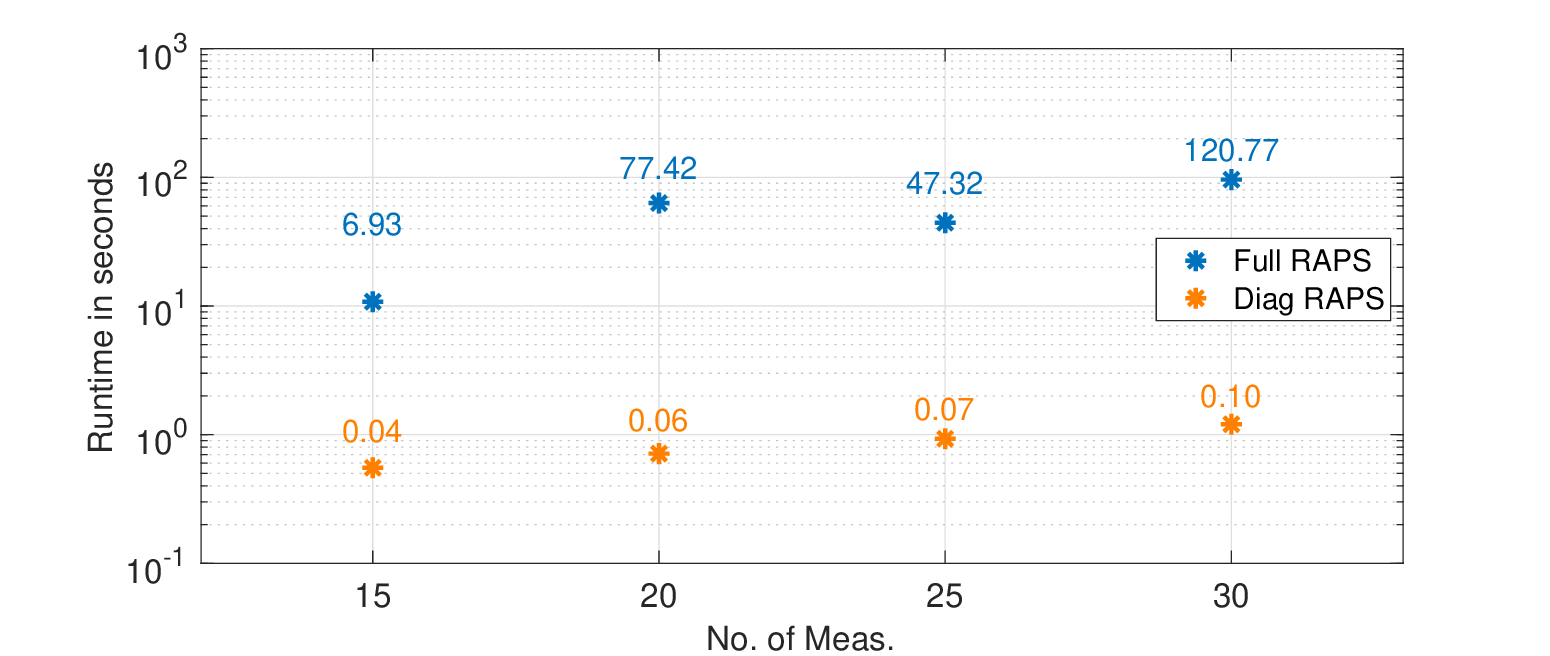}
	\caption{Runtime comparison between \ac{Full-RAPS} (blue) vs \ac{Diag-RAPS} (orange).  Each point represents the mean of the per epoch runtime for an experiment with the number of measurements specified along the horizontal axis. The values above the points represent the \ac{STD}. } 
	\label{fig:vs}
\end{figure}

As $m$ increases to 20, the \ac{Full-RAPS}  runtime extends beyond one minute per measurement epoch, rendering the evaluation of the complete dataset impractical. 
Consequently, the comparison focuses on a subset of 20 measurement epochs. 
Fig. \ref{fig:vs} compares the computational times for each method as a function of the number of measurements. 
The average computation time trends upward with the number of measurements.
\ac{Diag-RAPS} achieves a significantly lower computational cost, approximately 100 times faster than \ac{Full-RAPS}, and exhibits a reduced \ac{STD}. 
Based on these findings, only \ac{Diag-RAPS} was investigated for  state estimation performance comparisons.

\subsection{Estimator Implementations}
The time propagation  for all three estimators
 is 
 identical,
using the state and covariance propagation stated in eqns. \eqref{eqn:state_tm_update} and  \eqref{eqn:cov_tm_update}.
The computation of the parameters $\BoldF$ and $\BoldQ$ for the discrete-time model  discussed in Section 4.7 of \cite{farrell2008aided}  and derived in Appendix B of \cite{hu2024optimization}.

In this section, the number of measurements is  $m=50$.
Each estimator implements the  measurement update differently resulting in
a different  risk as quantified by a cost equivalent to eqn. \eqref{eqn:new_cost}. 
\begin{itemize}
	\item \ac{KF}: The standard \ac{KF} uses the $\Boldx$ that minimizes the MAP cost in eqn. \eqref{eqn:state_estimate}  for $\Boldb=\1$. 
	
	\item \ac{TD}: The \ac{KF} with \ac{TD} uses the $\Boldx$ that minimizes the MAP cost in eqn. \eqref{eqn:state_estimate}  with $\Boldb=\Boldb_{TD}$ (see eqn. \eqref{eqn:threshold_test}).

	\item \ac{RAPS}:  \ac{Diag-RAPS} uses the $\Boldx$ that solves Problem
	\ref{eqn:RAPS_diag_convex}.
\end{itemize}

\subsection{Results}

\begin{figure}[tb]
	\centering								
	\includegraphics[width=\columnwidth]{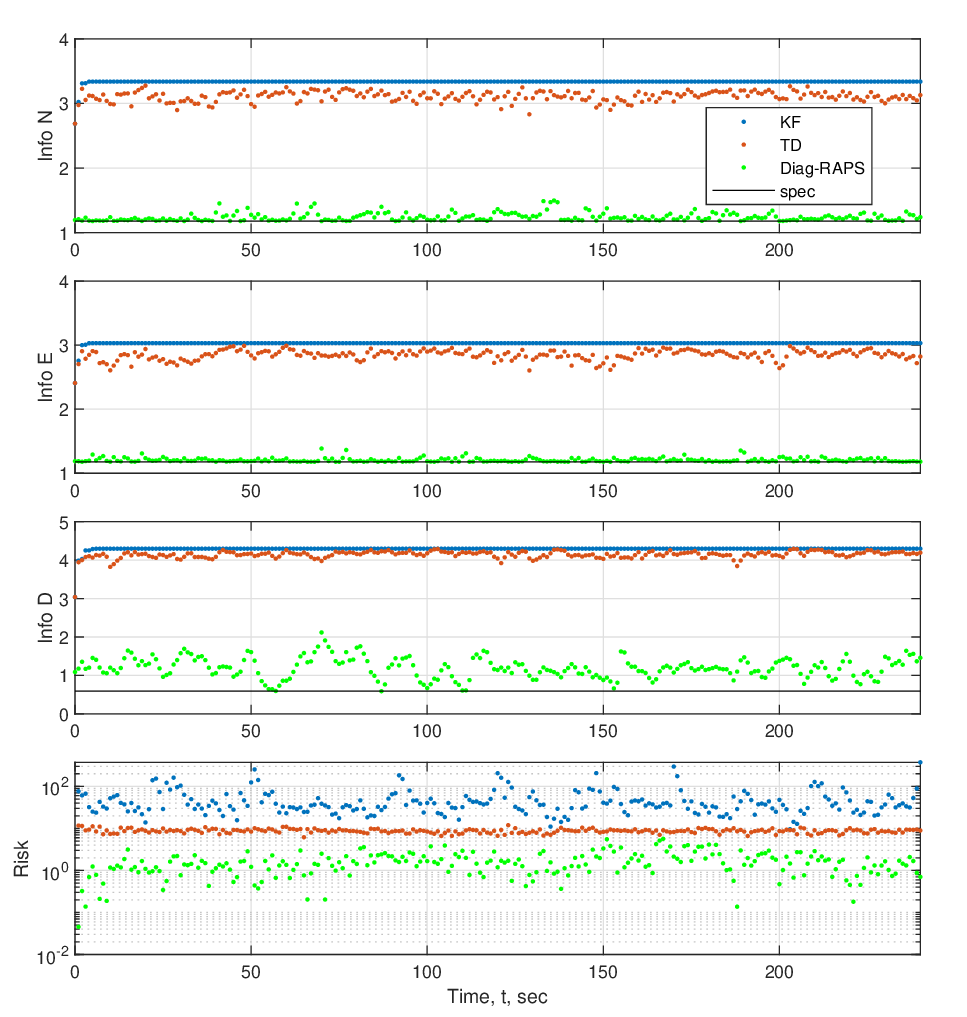}
	\caption{Simulation results comparing the square root of the posterior information and of the risk for Kalman filter (KF) in blue, Kalman filter with threshold detection (TD) in red, and \ac{Diag-RAPS} in green. The solid black line shows the corresponding square root of the diagonal element of $J_d$.}\label{fig:InfoRisk}
\end{figure}

\begin{figure}[tb]
	\centering
	\includegraphics[width=\columnwidth]{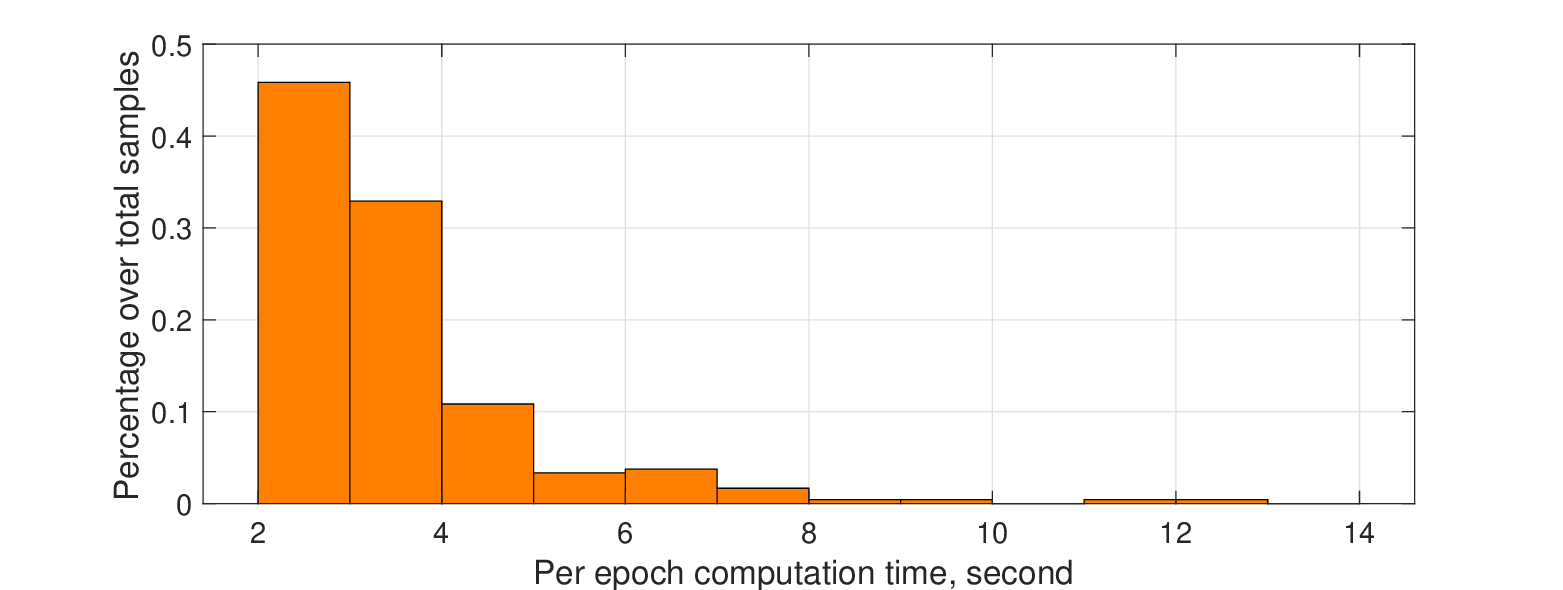}
	\caption{\ac{Diag-RAPS} computation time probability histogram.} \label{fig:compt}
\end{figure}

The estimation accuracies for the three estimators are shown in Fig. \ref{fig:InfoRisk}. 
The top three subplots show the square root of diagonal elements of the  posterior information matrix, which correspond to the north, east, and down position states.
The bottom subplot shows the square root of the risk incurred by each estimator.

Each plot of the square root of the posterior information shows:
the information achieved by the \ac{KF} as blue dots;
the information achieved by the \ac{TD} as red dots;
the information achieved by \ac{Diag-RAPS} as green dots;
the specified level of performance (i.e., square root of the corresponding component of $\BoldJ_d$) as a solid black line.
The same color scheme is used for the plot of the square root of the risk.
\black

The \ac{KF} graphs always achieves the highest (predicted) information for all components, but also the highest risk, because it uses all measurements. 
The \ac{TD} graphs show that, by ignoring those measurements whose normalized residual fails the threshold test, it is able to reduce the risk. 
The trade-off is that less (predicted) information is extracted from the measurements.
The \ac{Diag-RAPS} graphs show that it consistently maintains the lowest risk while always achieving the specified performance levels.
It is critical to understand that a high-level of predicted information is not likely to be achieved when the outlier risk is also high.

In terms of computational efficiency, both \ac{KF} and \ac{TD} methods require only a few milliseconds per measurement epoch, as they do not involve real-time optimization problems. 
Fig. \ref{fig:compt} presents a histogram showing the percentage of computation times (per measurement epoch) within each bin for \ac{Diag-RAPS} over 240 epochs, showing that it typically resolves within a few seconds.

\section{Conclusion and Future Research} \label{sect:conclusion}
	The main contributions of this article are the presentation and demonstration of a convex reformulation of the \ac{RAPS} state estimation problem. 
The article applied this  novel  approach to two variations of \ac{RAPS}  to attain their globally optimal solution. 
The \ac{Full-RAPS} approach imposes constraints on the full information matrix, whereas the \ac{Diag-RAPS} approach constrains the diagonal elements of the information matrix. Both of them yield a mixed-integer convex programming problem that can be resolved using existing software optimization tools.
Our simulation evaluations  demonstrate  that \ac{Diag-RAPS} is approximately 100 times faster than \ac{Full-RAPS}, making it a more practical choice for real-world applications.
The state estimation performance of \ac{Diag-RAPS} is 
 benchmarked against  a standard \ac{KF} using all measurements and the \ac{TD} approach based on checking normalized residual magnitudes against a threshold. 
The convex reformulation has proven effective, with simulation outcomes indicating that the RAPS solution meets the defined performance specifications while minimizing risk.

Regarding future research, this paper has concentrated on the solvability aspects of the \ac{RAPS} problem.
A very interesting next step is to study  the actual positioning performance of \ac{Diag-RAPS} using real-world data. 
This necessitates reducing computation costs to facilitate real-time application feasibility \cite{ding-24-detection}. 
On alternative approach to drastically improving computation time is based on the idea that the globally optimal measurement selection is not required; instead, what is required is a feasible solution that is not high  risk\cite{hu2024rapsppp}.

\section{Acknowledgment}
	The authors gratefully acknowledge the UCR KA Endowment and the US DOT CARNATIONS Center, each of which has partially funded this research. 
The ideas reported herein, and any errors or omissions, are the responsibility of the authors and do not reflect the opinions of the sponsors.

\bibliographystyle{biblio/IEEEtran}
\bibliography{biblio/IEEEabrv,biblio/References.bib}


\end{document}